\documentclass[preprintnumbers,eqsecnum,aps,prd,epsf,showpacs,nofootinbib]{revtex4} 
\usepackage{latexsym,amsmath,amssymb}
\usepackage[dvips]
{graphicx,color}
\usepackage{dcolumn,bm}    
\usepackage[utf8]{inputenc}
\newcommand{\ma}[1]{{\mathrm{#1}}}

\newcommand{\pa}{{\partial}}

\newcommand{\dd}{{\rm d}}

\newcommand{\wh}[1]{{\widehat{#1}}}
\newcommand{\wt}[1]{{\widetilde{#1}}}

\begin{document}
\thispagestyle{empty}


\title{Gravitational scalar-tensor theory}

\author{Atsushi Naruko}
 \email[Email : ]{naruko_at_th.phys.titech.ac.jp}
\author{Daisuke Yoshida}
 \email[Email : ]{yoshida_at_th.phys.titech.ac.jp}
\affiliation{Department of Physics, Tokyo Institute of Technology,
 Tokyo 152-8551, Japan}
\author{Shinji Mukohyama}
 \email[Email : ]{shinji.mukohyama_at_yukawa.kyoto-u.ac.jp}
 \affiliation{Yukawa Institute for Theoretical Physics, 
 Kyoto University, Kyoto 606-8502, Japan}
 \affiliation{Kavli Institute for the Physics and Mathematics of the Universe (WPI),
 The University of Tokyo, Chiba 277-8583, Japan}
\date{\today}

\begin{abstract}
We consider a new form of theories of gravity in which the action is written in terms of the Ricci scalar and its first and second derivatives. Despite the higher derivative nature of the action, the theory is free from ghost under an appropriate choice of the functional form of the Lagrangian. This model possesses $2+2$ physical degrees of freedom, namely $2$ scalar degrees and $2$ tensor degrees. We exhaust all such theories with the Lagrangian of the form $f(R, (\nabla R)^2, \Box R)$, where $R$ is the Ricci scalar, and then show some examples beyond this ansatz. In course of analysis, we prove the equivalence between these examples and a subclass of generalized bi-Galileon theories. 
\end{abstract}

\pacs{
} 

\preprint{YITP-15-117, IPMU 15-0211}

\maketitle

\section{Introduction}

Two major mysteries in modern cosmology are inflation and dark energy. Scalar-tensor theory is often invoked to describe those two phenomena (see e.g. for reviews \cite{Clifton:2011jh,Koyama:2015vza}). A conventional definition of scalar-tensor theory will be given as follows\,: theory of a scalar field (or fields) coupled to gravity.

On the other hand, there is an interesting mathematical tool, namely
 conformal (Weyl) transformation, which is the transformation of a metric
 where the metric is scaled by a spacetime dependent factor, $\Omega (\bm{x})$
 as $g_{\mu \nu} \to \Omega^2 ({\bm x}) \, g_{\mu \nu}$.
It is widely known that, under a conformal transformation,
 a canonical scalar field theory coupled to the Einstein gravity
 is mapped to a so-called $f (R)$ theory
 where $R$ represents the Ricci scalar (for a review of $f (R)$ theory see e.g. 
\cite{DeFelice:2010aj}).
Under this transformation, the degree of freedom of a scalar field
 in the original theory is replaced by the functional degree of freedom of
 $f(R)$. 
This is the reason why $f(R)$ theory is also dubbed as a scalar-tensor theory
 while no apparent scalar field exist in the action at a first glance.

There were significant recent developments of scalar-tensor theories in gravity,
 for example, the re-discovery of Horndeski's theory
 \cite{Horndeski:1974wa,Deffayet:2011gz,Kobayashi:2011nu}
 which is the most general single scalar field theory with
 second order differential equations
 with respect to metric and the scalar field
 where only $2+1$ degrees of freedom can propagate.
This theory is further extended keeping the same number of physical degrees of
 freedom as standard scalar-tensor theories or the Horndeski's theory
 while apparent higher derivative terms appear in the action
 in general \cite{Gleyzes:2014dya, Gao:2014soa,Lin:2014jga,Deffayet:2015qwa,Langlois:2015cwa,Langlois:2015skt}
(for further investigations of those theories see also \cite{Gleyzes:2014qga,Gleyzes:2014rba,Kobayashi:2014ida,Koyama:2015oma,Saito:2015fza,DeFelice:2015isa,Tsujikawa:2015mga,Tsujikawa:2015upa,Sakstein:2015zoa}).
Then a natural question will arise\,:
``Can we reformulate those generalized scalar-tensor theories in terms of the metric and its derivatives only, without using a scalar field in the action?'' 

To investigate a class of such theories, we shall consider a theory of gravity
 whose action is composed of the Ricci scalar and its derivatives\,:
 \begin{align}
 S = \frac{1}{\kappa^2}
 \int \dd^4 x \sqrt{- g} \,
 f \Bigl( R \,, (\nabla R)^2 \,, \Box R \Bigr) \,,
 \label{eqn:G-action}
 \end{align}
where $ (\nabla R)^2 = g^{\mu \nu} \nabla_\mu R \, \nabla_\nu R$ 
 and $\kappa^2 = 8 \pi G$ represents the gravitational constant
 while we suppress it in other equations for brevity unless otherwise stated.
Usually those higher derivative terms
 associated with derivatives of the Ricci scalar 
 introduce new and ghosty degrees of freedom 
 through its higher derivative nature, which is commonly called
 ``Ostrogradski instability'' \cite{Ostrogradsky:1850,Pais:1950za,Motohashi:2014opa}.
However as we show in this paper, this model does not suffer from such instabilities.Indeed there exists a theory in which only healthy $2+2$ dynamical degrees of freedom can propagate without ghost nor Ostrogradski instabilities under a particular and appropriate choice of the functional form of
 the Lagrangian. We then further extend the theory by including higher derivatives of the Ricci scalar. 
These theories can be understood as gravitational counterparts (dual) of
 standard multi-scalar-tensor theories composed of the metric and scalar fields. 
In fact this is the reason why we dub this theory as ``Gravitational scalar-tensor theory"
  because it is constructed only in terms of gravitational language, namely metric and its derivatives.
With our best knowledge, this kind of derivative terms have not been intensively investigated so far in the literature \cite{Wands:1993uu,Rodrigues:2011zi}. And hence this study can potentially open up a new direction in the study of gravity theory.

\section{New gravitational scalar-tensor theory}
\label{sec:simplenewtheory}

Let us investigate the nature of a gravity theory described by the action of the form (\ref{eqn:G-action}). In the case of the standard $f (R)$ gravity, we invoke an equivalent Lagrangian linear in $R$ instead of a non-linear function of $R$ by introducing a Lagrange multiplier and the associated auxiliary field\,:
 \begin{align}
 \int \dd^4 x \, \sqrt{- g} \, f (R)
 = \int \dd^4 x \, \sqrt{- g} \, \Bigl[ f (\phi)
 - \lambda (\phi - R) \Bigr] \,.  
 \end{align}
In the presence of derivatives of $R$, it is not trivial whether
 we can simultaneously replace derivatives of $R$ with
 the corresponding derivatives of $\phi$.
Interestingly we can do so as it is verified here.
In the case of $f \bigl( R \,, (\nabla R)^2 \,, \Box R \bigr)$,
first let us introduce a set of Lagrange multipliers $(\wt{\lambda} \,, \wt{\Lambda}_1 \,, \wt{\Lambda}_2)$ and the associated auxiliary fields $(\phi \,, X \,, B)$ to reduce the order of derivatives\,:
 \begin{align}
 S &= \int \dd^4 x \, \sqrt{- g} \, \Bigl[ f (\phi \,, X \,, B)
 - \wt{\lambda} \bigl( \phi - R \bigr) 
 - \wt{\Lambda}_1 \bigl( X - (\nabla R)^2 \bigr)
 - \wt{\Lambda}_2 \bigl( B - \Box R \bigr) \Bigr] \,. 
 \end{align}
Equivalently, one can rewrite this action as
 \begin{align}
 S &= \int \dd^4 x \, \sqrt{- g} \, \Bigl[ f (\phi \,, X \,, B)
 - \lambda \bigl( \phi - R \bigr) 
 - \Lambda_1 \bigl( X - (\nabla \phi)^2 \bigr)
 - \Lambda_2 \bigl( B - \Box \phi \bigr) \Bigr] \,,
 \end{align}
where $\lambda=\wt{\lambda}-\nabla^{\mu}[\wt{\Lambda}_1\nabla_{\mu}(\phi+R)]+\Box\wt{\Lambda}_2$, $\Lambda_1=\wt{\Lambda}_1$, $\Lambda_2=\wt{\Lambda}_2$. This replacement is justified since the transformation from $(\wt{\lambda} \,, \wt{\Lambda}_1 \,, \wt{\Lambda}_2)$ to $(\lambda\,, \Lambda_1\,,\Lambda_2)$ is regular and invertible.
Since the action does not include any derivative terms of $\Lambda_1$ and $\Lambda_2$, variation of the action with respect to those variables yield constraint equations rather than dynamical equations of motion, which can be plugged back into the action without changing the nature of the theory\,:
 \begin{align}
 S &= \int \dd^4 x \, \sqrt{- g} \, \Bigl[
 f \Bigl( \phi \,, (\nabla \phi)^2 \,, \Box \phi \Bigr)
 - \lambda \bigl( \phi - R \bigr) \Bigr] \,. 
 \label{eqn:action}
 \end{align}
This verifies that we can indeed replace all the derivatives of $R$ with
 those of $\phi$ under the replacement of $R$ by $\phi$
 and the introduction of the Lagrange multiplier.
It should be noted that due to the presence of derivative terms of $\phi$,
 the extremization of the action with respect to $\phi$ gives 
 a dynamical equation of $\phi$ rather than a constraint equation,
 which cannot be plugged back into the action. 
And hence now $\lambda$ should be treated as one of dynamical fields.
This is in sharp contrast to the case of $f(R)$ where $\lambda$ can be determined by $f$, that is $\lambda = \dd f(\phi)/\dd \phi$.

Now higher derivative terms except for $\lambda R$ only come from $\Box \phi$.
In order to reduce the order of derivatives, it is convenient to (re)introduce
 a Lagrange multiplier and the associated auxiliary field as
 \begin{align}
 S &= \int \dd^4 x \, \sqrt{- g} \, \Bigl[
 f \Bigl( \phi \,, (\nabla \phi)^2 \,, B \Bigr)
 - \lambda \bigl( \phi - R \bigr) 
 - \Lambda \bigl( B - \Box \phi \bigr) \Bigr] \,. 
 \end{align}
Given the fact that the action does not depend on derivative of $B$,
 variation of the action with respect to $B$ gives a constraint equation,
\begin{align}
 \Lambda = f_B \,. \label{eqn:Lambda}
\end{align}
Hereafter, subscript $B$ to $f$ denotes partial derivative with respect to it. The constraint (\ref{eqn:Lambda}) enables us to eliminate $\Lambda$ in the action by plugging it back into the action, provided that
\begin{equation}
 f_{BB} \ne 0. \label{eqn:nonvanishing-hessian}
\end{equation}
Under (and only under) this condition, substituting (\ref{eqn:Lambda}) to the action and then variating the resulting action with respect to $B$ leads to the original relation $B=\Box\phi$. We thus first consider the case with (\ref{eqn:nonvanishing-hessian}). 

The action except for the $\lambda R$ term reduces to a first order form after integration by parts, 
 \begin{align}
 S &= \int \dd^4 x \, \sqrt{- g} \, \Bigl( 
 \lambda R - g^{\mu \nu} \pa_\mu f_B \pa_\nu \phi
 + f - f_B B - \lambda \phi \Bigr) \,.   
 \end{align}
If $f_{BB}\ne 0$ then we can define a new variable $\varphi$ as 
\begin{equation}
 \varphi \equiv f_B\,, \label{eqn:def-varphi}
\end{equation}
and consider ($g_{\mu \nu}$, $\lambda$, $\phi$, $\varphi$) as basic variables. This is justified since the transformation from ($g_{\mu \nu}$, $\lambda$, $\phi$, $B$) to ($g_{\mu \nu}$, $\lambda$, $\phi$, $\varphi$) is locally invertible because of non-vanishing $f_{BB}$.
Moreover it is convenient to move to the Einstein frame via a conformal transformation. 
Under the assumption that tensor modes have positive kinetic terms, i.e. 
\begin{equation}
 \lambda >0\,, \label{eqn:noghost-tensor}
\end{equation}
after a conformal transformation such that $g_{\mu \nu} = \Omega^2 \, \wh{g}_{\mu \nu}$ with $\Omega^{-2} = 2 \lambda$ the action can be cast into the form of the action of multi-scalar fields coupled to the Einstein gravity\,:
\begin{equation}
 S = \int \dd^4 x \, \sqrt{- \wh{g}} \, 
  \left\{ \frac{1}{2} \wh{R}
   - \frac{1}{2}\wh{g}^{\mu \nu} 
   \left(
    \partial_{\mu}\chi\partial_{\nu}\chi
    + e^{-\sqrt{\frac{2}{3}} \chi}\partial_{\mu}\varphi\partial_{\nu}\phi
   \right)
  - \frac{1}{4} \left[ e^{-\sqrt{\frac{2}{3}}\chi} \phi
  + e^{-2\sqrt{\frac{2}{3}}\chi} \Bigl( \varphi \, B \bigl(\phi,(\wh{\nabla} \phi)^2,\varphi \bigr) - f \Bigr) \right]
  \right\} \,. \label{eqn:action-varphi}
 \end{equation}
Here, we have introduced a new variable $\chi$ via
\begin{equation}
 \lambda \equiv e^{\sqrt{\frac{2}{3}} \chi} \,, 
 \label{def:chi}
\end{equation}
which has a canonically normalized kinetic term. It is easy to see that the kinetic matrix has one negative eigenvalue, meaning the existence of a ghost. This is a consequence of the cross term $\wh{\nabla}_\mu \varphi \wh{\nabla}^\mu \phi$ in the absence of $\wh{\nabla}_\mu \varphi \wh{\nabla}^\mu \varphi$. On the other hand, $\chi$ has the standard kinetic term $\wh{\nabla}_\mu \chi \wh{\nabla}^\mu \chi$ and no cross term. Because of this structure, the $3\times 3$ kinetic matrix has two positive eigenvalues and one negative eigenvalue. Therefore, irrespective of the functional form of $f$, there is one ghosty mode owing to this structure of the Lagrangian. 

Here comments are in order. Since now $\lambda$ and correspondingly $\chi$ are dynamical fields, strictly speaking, it is not a priori guaranteed that (\ref{eqn:noghost-tensor}) is satisfied all the time through the evolution of the system. However, in fact, if the condition is initially satisfied, it always holds at least in the Einstein frame. Since now $\lambda$ is one of dynamical fields, one can always choose initial conditions such that this condition is satisfied. Then, as is clear from (\ref{def:chi}), $\lambda = 0$ corresponds to the infinity in the moduli space where $\chi$ approaches negative infinity and hence the condition (\ref{eqn:noghost-tensor}) will be never violated at least with the finite lapse of time (in the Einstein frame).

Next let us investigate an interesting case with $f_{BB}=0$. In this case, since $B$ or correspondingly $\Box R$ linearly enters in the Lagrangian, we can rewrite $f$ in a simpler manner as
\begin{equation}
 f \Bigl( R \,, (\nabla R)^2 \,, \Box R \Bigr)
  = {\cal K} \Bigl( R \,, (\nabla R)^2 \Bigr)
  + {\cal G} \Bigl( R \,, (\nabla R)^2 \Bigr) \Box R \,.
  \label{eqn:fBB=0case}
\end{equation}
Here it should be noted that if ${\cal G}$ is a function of $R$ solely,
 the second term can be absorbed into
 the first one after integration by parts
 \begin{align}
 {\cal K} \Bigl( R \,, (\nabla R)^2 \Bigr) 
 + {\cal G} (R) \Box R 
 = \wt{{\cal K}} \Bigl( R \,, (\nabla R)^2 \Bigr) 
 + (\ma{tot. ~ der.})\,, \quad
  \wt{{\cal K}} \Bigl( R \,, (\nabla R)^2 \Bigr) 
  \equiv 
  {\cal K} \Bigl( R \,, (\nabla R)^2 \Bigr) 
 - {\cal G}'(R) (\nabla R)^2\,.
\label{case-K}
 \end{align}
Hence this case is equivalent to the case with ${\cal G}=0$. In any case, irrespective of whether ${\cal G}$ depends on $(\nabla R)^2$ or not, after following the same steps as before, one obtains the following action\,:
 \begin{align}
 S &= \int \dd^4 x \, \sqrt{- \wh{g}} \, \left[
 \frac{1}{2} \wh{R}  
 - \frac{1}{2} \wh{g}^{\mu \nu} \left( \pa_\mu \chi \pa_\nu \chi  
 + \sqrt{\frac{2}{3}} e^{- \sqrt{\frac{2}{3}} \chi} \, {\cal G}
 \, \pa_\mu \chi \pa_\nu \phi \right)
 + \frac{1}{2} e^{- \sqrt{\frac{2}{3}} \chi} \, {\cal G} \,
 \widehat{\overset{\,}{\Box}} \phi
 + \frac{1}{4}e^{-2\sqrt{\frac{2}{3}}\chi} \, {\cal K}
  - \frac{1}{4}e^{-\sqrt{\frac{2}{3}}\chi} \, \phi
  \right] \,, \label{eqn:S-fBB=0}
 \end{align}
where it is understood that
 \begin{align}
{\cal K} = 
  {\cal K} \Bigl(\phi, \, 
  2e^{\sqrt{\frac{2}{3}}\chi} \, \wh{g}^{\, \mu\nu}\partial_{\mu}\phi \, \partial_{\nu}\phi \Bigr)\,,
  \quad
{\cal G} = 
  {\cal G} \Bigl( \phi, \,
  2e^{\sqrt{\frac{2}{3}}\chi} \, \wh{g}^{\, \mu\nu} \, \partial_{\mu}\phi \, \partial_{\nu}\phi \Bigr)\,.
 \end{align}
One of the most remarkable differences from the previous case is that
 now both $\phi$ and $\chi$ fields can have healthy kinetic terms
 at least by properly choosing the functional form of ${\cal K}$ and ${\cal G}$.
Although there is a cross term, $\wh{\nabla}_\mu \chi \wh{\nabla}^\mu \phi$, 
in the presence of non-vanishing ${\cal G}$, this term does not necessarily cause a problem since both ${\cal K}$ and ${\cal G}$ depend on $\wh{\nabla}_\mu \phi \wh{\nabla}^\mu \phi$. Due to this crucial difference in the structure of the Lagrangian, 
 this case will end up with a healthy theory without Ostrogradski nor ghost 
 instabilities, which is actually confirmed by perturbative analysis
 around some background e.g. cosmological one.
In the absence of ${\cal G}$ as well as in the case of (\ref{case-K}),
 the cross term, $\wh{\nabla}_\mu \chi \wh{\nabla}^\mu \phi$ disappears 
 and hence this case will also be healthy since
 both $\phi$ and $\chi$ fields can have healthy kinetic terms
 under an appropriate choice of the functional form of ${\cal K}$.

As an example, let us consider the simple case with
 ${\cal K} = - (\nabla R)^2/2$ and ${\cal G} = 0$.
In this case the equivalent action (\ref{eqn:S-fBB=0}) is reduced to
 \begin{align}
 S &= \int \dd^4 x \, \sqrt{- \wh{g}} \, \left[
 \frac{1}{2 \, \kappa^2} \wh{R}  
 - \frac{1}{2} \wh{g}^{\, \mu \nu} \pa_\mu \wh{\chi} \, \pa_\nu \wh{\chi} 
 - \frac{1}{4} e^{- \sqrt{\frac{2}{3}} \kappa \wh{\chi}} \,
  \wh{g}^{\, \mu \nu} \pa_\mu \wh{\phi} \, \pa_\nu \wh{\phi} 
  - \frac{1}{4 \kappa} e^{-\sqrt{\frac{2}{3}} \kappa \wh{\chi}} \, \wh{\phi} \,
  \right] \,,
 \end{align}
 where we have recovered the gravitational constant for clarity and
 $\wh{\chi}$ and $\wh{\phi}$ are introduced by
 \begin{align}
 \chi \equiv \kappa \, \wh{\chi} \,, \qquad
 \phi \equiv \kappa \, \wh{\phi} \,.  
 \end{align}
This system is simply the Einstein gravity with the two minimally-coupled scalar fields $\chi$ and $\phi$, whose $2\times 2$ kinetic matrix is manifestly positive definite. Therefore, it is obvious that there is no ghost in the system, at least at time and length scales sufficiently shorter than the curvature radius~\cite{GMS}.

\section{Further extensions}

In the previous section we have investigated theories whose action is of the form (\ref{eqn:G-action}) and showed that the theory can be free from ghost only if $f$ is of the form (\ref{eqn:fBB=0case}). This exhausts all possible theories with $2+2$ physical degrees of freedom without ghost if we start with the ansatz (\ref{eqn:G-action}) for the action. In this section we shall further extend the theory so as to include more higher derivative terms. We shall not provide a systematic search. Instead, we shall show some explicit examples of such theories. 

In order to keep the same number of degrees of freedom as in the last section, we implement the structure of the generalized Galileon theory, or Horndeski's theory. This suggest that the $(\Box R)^2$ term be accompanied by the term $(\nabla_\mu \nabla_\nu R)^2 = (\nabla_\mu \nabla_\nu R) (\nabla^\mu \nabla^\nu R)$. Then in terms of three arbitrary functions of $R$ and $(\nabla R)^2$, $({\cal K} \,, {\cal G} \,, {\cal Q})$, let us consider the action of the form $S=\int \dd^4x\sqrt{-g}\, f$, where
\begin{equation}
 f 
  = {\cal K} \Bigl( R \,, (\nabla R)^2 \Bigr) 
 + {\cal G} \Bigl( R \,, (\nabla R)^2 \Bigr) \Box R 
 + {\cal Q} \Bigl( R \,, (\nabla R)^2 \Bigr) \, R  
 - 2 \frac{\pa {\cal Q}}{\pa (\nabla R)^2}
 \Bigl( R \,, (\nabla R)^2 \Bigr) 
 \Bigl[ (\Box R)^2 - (\nabla_\mu \nabla_\nu R)^2 \Bigr]\,.
\label{quartic}
\end{equation}
By introducing a Lagrange multiplier and an auxiliary field to partly replace $R$, the action is rewritten as
\begin{align}
 S = \int \dd^4 x \sqrt{- g} \, \left\{
 {\cal K} (\phi \,, X) + {\cal G} (\phi \,, X) \Box \phi
 + {\cal Q} (\phi \,, X) \, R
 - 2 \frac{\pa {\cal Q}}{\pa X} (\phi \,, X) 
 \Bigl[ (\Box \phi)^2 - (\nabla \nabla \phi)^2 \Bigr]   
 - \lambda (\phi - R) \,\right\}\,,
\end{align}
where $X = (\nabla \phi)^2$. This is a special case of the generalized bi-Galileons~\cite{Padilla:2012dx,Kobayashi:2013ina,Ohashi:2015fma}. Therefore, the number of physical degrees of freedom in this theory is $2+2$. 

Next let us add a term corresponding to the so-called quintic Horndeski term. For this purpose it is convenient to rewrite $G_{\mu\nu}\nabla^{\mu}\nabla^{\nu}R$ in terms of $R$ and its derivatives, where $G_{\mu\nu}$ is the Einstein tensor. First by the definition of the Riemann tensor we have
\begin{align}
 (\nabla_\beta \nabla_\gamma - \nabla_\gamma \nabla_\beta) \nabla_\alpha \psi
 = R^\delta_{\ \alpha \beta \gamma} \nabla_\delta \psi \,,
\end{align}
for an arbitrary scalar $\psi$. By contracting this with inverse metric $g^{\alpha \gamma}$ and taking the divergence of it, one obtains
\begin{equation}
 \nabla^\alpha (\nabla_\alpha \Box - \Box \nabla_\alpha) \psi
 = R_{\alpha \beta} \nabla^\alpha \nabla^\beta \psi 
 + (\nabla^\alpha R_{\alpha \beta}) \nabla^\beta \psi 
 = R_{\alpha \beta} \nabla^\alpha \nabla^\beta \psi 
 + \frac{1}{2} \nabla_\alpha R \, \nabla^\alpha \psi \,,
\end{equation} 
where in the last equality we have utilized the Bianchi identity. Finally, by replacing $\psi$ with $R$ in the above equation, one arrives at the identity of the form
 \begin{align}
 G_{\mu \nu} \nabla^\mu \nabla^\nu R
 = \nabla^\alpha (\nabla_\alpha \Box - \Box \nabla_\alpha) R
 -\frac{1}{2}(\nabla R)^2 - \frac{1}{2} R \, \Box R \,.
 \end{align}
Hence one can add the following term to (\ref{quartic}). 
\begin{align}
 & {\cal C} \Bigl( R \,, (\nabla R)^2 \Bigr) \, \left[
 \nabla^\alpha (\nabla_\alpha \Box - \Box \nabla_\alpha) R
 -\frac{1}{2}(\nabla R)^2 - \frac{1}{2} R \, \Box R \right] \notag\\ 
 & \qquad
 + \frac{1}{3} \frac{\pa \, {\cal C}}{\pa (\nabla R)^2}
 \Bigl( R \,, (\nabla R)^2 \Bigr) \, \Bigl[ (\Box R)^3
 - 3 (\Box R) (\nabla_\mu \nabla_\nu R)^2 + 2 (\nabla_\mu \nabla_\nu R)^3
 \Bigr] \,,
\end{align}
where $(\nabla_\mu \nabla_\nu R)^3 = (\nabla_\mu \nabla_\nu R) (\nabla^\nu \nabla^\rho R) (\nabla_\rho \nabla^\mu R)$. By introducing the auxiliary field $\phi$ and the Lagrange multiplier $\lambda$, the action is rewritten as
\begin{eqnarray}
 S & = & \int \dd^4 x \sqrt{- g} \, \left\{
 {\cal K} (\phi \,, X) + {\cal G} (\phi \,, X) \Box \phi
 + {\cal Q} (\phi \,, X) \, R
 - 2 \frac{\pa {\cal Q}}{\pa X} (\phi \,, X) 
 \Bigl[ (\Box \phi)^2 - (\nabla \nabla \phi)^2 \Bigr]
 - \lambda (\phi - R)
 \right.
 \nonumber\\
 & & \left.
+ \, {\cal C}(\phi\,, X)\, \left[
 \nabla^\alpha (\nabla_\alpha \Box - \Box \nabla_\alpha) \phi
 -\frac{1}{2}(\nabla \phi)^2 - \frac{1}{2} \phi \, \Box \phi \right]
 + \frac{1}{3} \frac{\pa \, {\cal C}}{\pa X}(\phi\,, X) \, \Bigl[ (\Box \phi)^3
 - 3 (\Box \phi) (\nabla_\mu \nabla_\nu \phi)^2 + 2 (\nabla_\mu \nabla_\nu \phi)^3
 \Bigr] \,\right\}\,,\nonumber\\
\end{eqnarray}
where $(\nabla_\mu \nabla_\nu \phi)^3 = (\nabla_\mu \nabla_\nu \phi) (\nabla^\nu \nabla^\rho \phi) (\nabla_\rho \nabla^\mu \phi)$. Again, this is a special case of the generalized bi-Galileons~\cite{Padilla:2012dx,Kobayashi:2013ina,Ohashi:2015fma} and thus the number of physical degrees of freedom in this theory is $2+2$. 

A multi-field extension of the yet another generalizations performed in \cite{Gleyzes:2014dya,Gao:2014soa} can be also implemented in a similar way.

\section{Summary and Discussion}

We have considered a new form of scalar-tensor theory where the action is composed of Ricci scalar and its first and second derivatives, which is dubbed as ``Gravitational scalar-tensor theory"
   because it can be written only in terms of gravitational language, that is metric and its derivatives.
Surprisingly, despite the higher derivative nature of the Lagrangian, the theory is healthy in the sense that there is no ghost nor Ostrogradski instabilities under an appropriate choice of the functional form of the Lagrangian. This model only possesses $2+2$ degrees of freedom, namely 2 for scalar degrees of freedom and 2 for tensors. We have also extended the theory so as to include more higher derivative terms while keeping the same number of degrees of freedom as before and discussed the relation to the multi-field extension of Horndeski's theory. 

In the present paper, we have focused on the inclusion of the Ricci scalar and its derivatives only. It will be also interesting to include derivatives of Riemann tensor to extend the so-called $f(\ma{Riemann})$ theory \cite{Deruelle:2009zk}.

Finally, let us discuss about matter, especially its coupling to gravity. 
While we have only focused on the gravitational sector, matter sector must be
 introduced in reality.
 In principle matter field can couple to either the original metric or
 the conformally related metrics. 
However, since the action of the gravitational sector of the theory is written in terms of the curvature and its derivatives, we consider it natural to couple matter field to the original metric $g_{\mu\nu}$, i.e. the Jordan frame metric. 
In this case, the conformal transformation to the Einstein frame if it exists
 affects the matter action. 
In particular, the gravitational force is mediated by not only the Einstein frame metric but also the two scalar fields in general. 
Also, the behavior of the two scalar fields in the gravity sector may depend on the environment provided by the matter fields. 
Phenomenology of this modified gravity theory is thus worthwhile investigating in more details.

\acknowledgments
A.N. would like to thank Xian Gao, Misao Sasaki, Masahide Yamaguchi and
 Daisuke Yamauchi for fruitful discussions.
A.N. is grateful to the Institut Astrophysique de Paris 
 and Yukawa Institute for Theoretical Physics at Kyoto University for warm hospitality
 where this work was advanced.
We would also like to thank the Yukawa Institute for Theoretical Physics at Kyoto University since discussions during the YITP workshop YITP-W-15-16 on "JGRG25" were useful to complete this work.
This work was supported in part by the JSPS Research Fellowship
 for Young Scientists Nos. 263409 (A.N.) and 2611495 (D.Y.).
One of the authors (SM) was supported in part by Grant-in-Aid for Scientific Research 24540256 and the WPI Initiative, MEXT Japan. Part of his work has been done within the Labex ILP (reference ANR-10-LABX-63) part of the Idex SUPER, and received financial state aid managed by the Agence Nationale de la Recherche, as part of the programme
Investissements d'avenir under the reference ANR-11-IDEX-0004-02. He is thankful to colleagues at Institut Astrophysique de Paris, especially Jean-Philippe Uzan, for warm hospitality.


\end{document}